\documentclass[lettersize,journal]{IEEEtran}

\usepackage{amsmath,amsfonts}
\usepackage{algorithmic}
\usepackage{algorithm}
\usepackage{array}
\usepackage{textcomp}
\usepackage{stfloats}
\usepackage{url}
\usepackage{verbatim}
\usepackage{graphicx}
\usepackage{cite}
\usepackage{comment}
\usepackage{booktabs}
\usepackage{tabularx}
\usepackage{multirow}
\usepackage{balance}
\usepackage[hidelinks]{hyperref}

\usepackage[caption=false,font=footnotesize,labelfont=sf,textfont=sf]{subfig}

\usepackage{titlesec}
\titlespacing{\section}{0pt}{2.5pt}{3pt}
\titlespacing{\subsection}{0pt}{4pt}{2pt}
\titlespacing{\subsubsection}{0pt}{3pt}{1.5pt}

\begin{document}

\title{Risk-Based PV-Rich Distribution System Planning \\Using Generative AI}

\author{Habtemariam Aberie Kefale,~\IEEEmembership{Student Member,~IEEE,} Weijie Xia,~\IEEEmembership{Student Member,~IEEE,}
Nanda Kishor Panda,~\IEEEmembership{Student Member,~IEEE,} Peter P. Palensky,~\IEEEmembership{Senior Member,~IEEE}, Pedro P. Vergara,~\IEEEmembership{Senior Member,~IEEE}
\thanks{This work was supported by the Marie Skłodowska-Curie Actions and Delft University of Technology via the GROW project (grant 101126487).}
        
\thanks{Habtemariam Aberie Kefale, Weijie Xia, Peter Palensky, and Pedro P. Vergara are with the Intelligent Electrical Power Grids Group, Delft University of Technology, 2628 CD Delft, The Netherlands; Nanda Kishor Panda is with Enexis Netbeheer B.V., 5201 AW 's-Hertogenbosch, The Netherlands (e-mail: h.a.kefale@tudelft.nl, W.Xia@tudelft.nl, P.Palensky@tudelft.nl, P.P.VergaraBarrios@tudelft.nl, nanda.kishor.panda@enexis.nl).}
 }

\markboth{}%
{Shell \MakeLowercase{\textit{et al.}}: A Sample Article Using IEEEtran.cls for IEEE Journals}
\maketitle
\begin{abstract}
Hosting capacity (HC) assessment plays a critical role in distribution system planning under increasing penetration of distributed energy resources (DERs) and associated uncertainties in load and generation. However, conventional approaches often rely on deterministic worst-case evaluation, leading to overly conservative HC estimates. This paper introduces a risk-based framework for HC assessment that explicitly accounts for the frequency, intensity, and duration of voltage violations under uncertain operating conditions. A generative AI-based approach is employed to generate realistic, time-correlated load demand scenarios conditioned on projected energy consumption growth levels. These scenarios are then used to assess voltage violations and quantify their risk using probabilistic intensity, duration, and frequency (IDF) metrics. The results show that extreme-percentile (zero-risk) approaches significantly underestimate PV-HC by treating all violations equally, regardless of their likelihood or persistence. For instance, allowing a 5\% risk level increases HC by approximately 18\% for a 15~min violation duration. The proposed approach provides a practical tool for risk-informed distribution system planning under uncertainty. 
\end{abstract}

\section{Introduction}
\IEEEPARstart{G}{lobally}, the integration of distributed energy resources (DERs), particularly distributed photovoltaic (PV) systems, is rapidly increasing as part of the transition toward more sustainable and low-carbon energy systems \cite{DER}, \cite{DIPPENAAR2026102104}. For example, in the Netherlands, installed PV capacity exceeded 24~GW in 2023 and is projected to surpass 50~GW by 2030, representing an increase of more than 100\% within this decade \cite{NPE2023}.

Consequently, the increased penetration and uncertainty of DERs introduce several technical challenges in distribution systems \cite{DER_review}, resulting in greater complexity for distribution system operators (DSOs) in both system operation and planning. To support informed planning decisions while ensuring secure system operation, DSOs must estimate the maximum DER capacity that can be accommodated without violating technical constraints or requiring costly grid reinforcements. This maximum limit, commonly referred to as Hosting Capacity (HC), has become a key metric for modern distribution system planning \cite{10415382, 10811912}.

In PV-rich distribution systems, voltage magnitude constraints often play a dominant role in PV hosting capacity (PV-HC) assessments \cite{pedro_paper}. In addition to increasing PV deployment, energy consumption is also expected to grow due to electrification of the transport and heating sectors. The combined growth of PV penetration and energy consumption introduces additional uncertainty into future system conditions, creating new challenges for distribution system planning and motivating approaches to assess and quantify the risk of technical violations under uncertain operating conditions.

Several studies have investigated PV-HC assessment using different approaches, which can be broadly classified into deterministic and probabilistic methods. In deterministic methods, PV penetration is incrementally increased until one or more technical constraints are violated under predefined operating conditions. The work in \cite{determistic} proposed a voltage sensitivity–based analytical method to estimate PV-HC in distribution systems, while \cite{determisitc2} determined it through distribution feeder simulations by progressively increasing PV penetration until voltage or current limits are violated. While these methods are computationally efficient and suitable for preliminary assessments, they neglect the inherent variability and uncertainties associated with PV generation and load demand. Consequently, the resulting HC estimates tend to be overly conservative or unrealistic \cite{7913608}, thereby limiting their suitability for planning in today's distribution systems that incorporate diverse DER technologies.

To better account for uncertainty in DER integration and load demand, probabilistic HC assessment methods have been widely investigated. A probabilistic framework is proposed in \cite{8325320} to determine the maximum acceptable penetration of PV and wind generation, accounting for uncertainties in load demand, generation output, and location. In this work, HC is formulated as a nonlinear optimization problem, where uncertainty in load demand and DER (PV and wind) is represented through scenarios generated by Monte Carlo (MC) simulation. To assess PV-HC in unbalanced distribution systems under coordinated voltage regulation, the authors in \cite{9745040} proposed a two-stage optimization-based framework. The method first determines the optimal PV base capacity and then evaluates PV limits through time-series power flow analysis, while incorporating PV uncertainty using MC–based simulations. The work in \cite{ali2020probabilistic} proposed another probabilistic framework to assess and enhance PV-HC by modeling PV generation and load demand using Beta and Normal distributions, respectively. The approach jointly optimizes inverter oversizing together with active power curtailment and reactive power support, enabling higher PV integration while mitigating voltage deviations. A probabilistic evaluation framework is further presented in \cite{ma2024probabilistic} to assess PV-HC in unbalanced distribution systems. The method incorporates voltage magnitude and voltage unbalance constraints and introduces a probabilistic violation risk index to quantify HC. To mitigate the required computational time to run time-series simulations, a polynomial chaos-based Kriging surrogate model is used.

While probabilistic HC assessment methods provide a more realistic representation of system behavior under uncertainty, their effectiveness depends on the quality and representativeness of the scenarios used in the analysis. In many existing approaches, scenarios are generated using simplified statistical assumptions, which may not adequately capture the temporal correlations and variability observed in real system operation. Recent advances in generative AI algorithms offer new opportunities to generate realistic scenarios, helping to better account for risk in distribution system planning. Several works have incorporated risk-aware formulations into HC assessment. For instance, the work in \cite{risk_aware} developed a risk-aware operating region framework to characterize admissible PV operating conditions under irradiance variability, where profiles were generated using advanced Copula models. Authors in \cite{10040563} formulated HC using risk-sensitive constraints based on conditional value at risk (CVaR), enabling the regulation of constraint violations across uncertain operating scenarios. Similarly, the work in \cite{9416873} incorporated risk aversion into planning decisions by explicitly accounting for uncertainty in DER deployment and system operation. Although these works represent important steps toward integrating risk considerations into HC assessment, they primarily focus on optimization formulations or operating limits rather than providing planning-oriented metrics that explicitly characterize technical violations. The impact of such violations depends on \textit{how often} they occur (defined as Frequency), \textit{how severe} they are (defined as Intensity), and \textit{how long} they persist (defined as Duration), making these aspects essential for distribution system planning. Consequently, there is a need to translate probabilistic simulation results into interpretable and planning-relevant risk indicators that enable informed long-term planning decisions.

In this paper, we propose a risk-based HC analysis framework that enables DSOs to make well-informed planning decisions. The proposed framework evaluates under- and overvoltage conditions through probabilistic simulations driven by advanced generative AI algorithms and translates the resulting system behavior into planning-relevant reliability indicators over the planning horizon. The main contributions of this paper are summarized as follows:

\begin{itemize}

\item A comprehensive distribution system planning framework is proposed to support DSOs in assessing the technical impacts of PV integration and long-term energy consumption growth under uncertainty. The conditional flow-based deep generative model in \cite{flow_model} is employed to generate future load demand scenarios conditioned on projected energy consumption growth levels. This conditioning enables the controlled generation of realistic future consumption patterns aligned with planning assumptions, while preserving temporal correlations and annual energy characteristics across different user types (e.g., residential, industrial).

\item An Intensity–Duration–Frequency (IDF)-based representation and visualization is introduced to assess HC under combined energy consumption growth, PV penetration, and irradiance variability. The proposed IDF framework characterizes stochastic voltage magnitude violations in terms of their severity, persistence, and frequency, providing relevant risk metrics for distribution system planning analysis. This representation enables DSOs to interpret probabilistic simulation outcomes more effectively and identify PV integration levels while explicitly quantifying the associated operational risk.
\end{itemize}
\section{Proposed Framework}
\begin{figure}[t]
    \centering
    \includegraphics[width=0.95\linewidth]{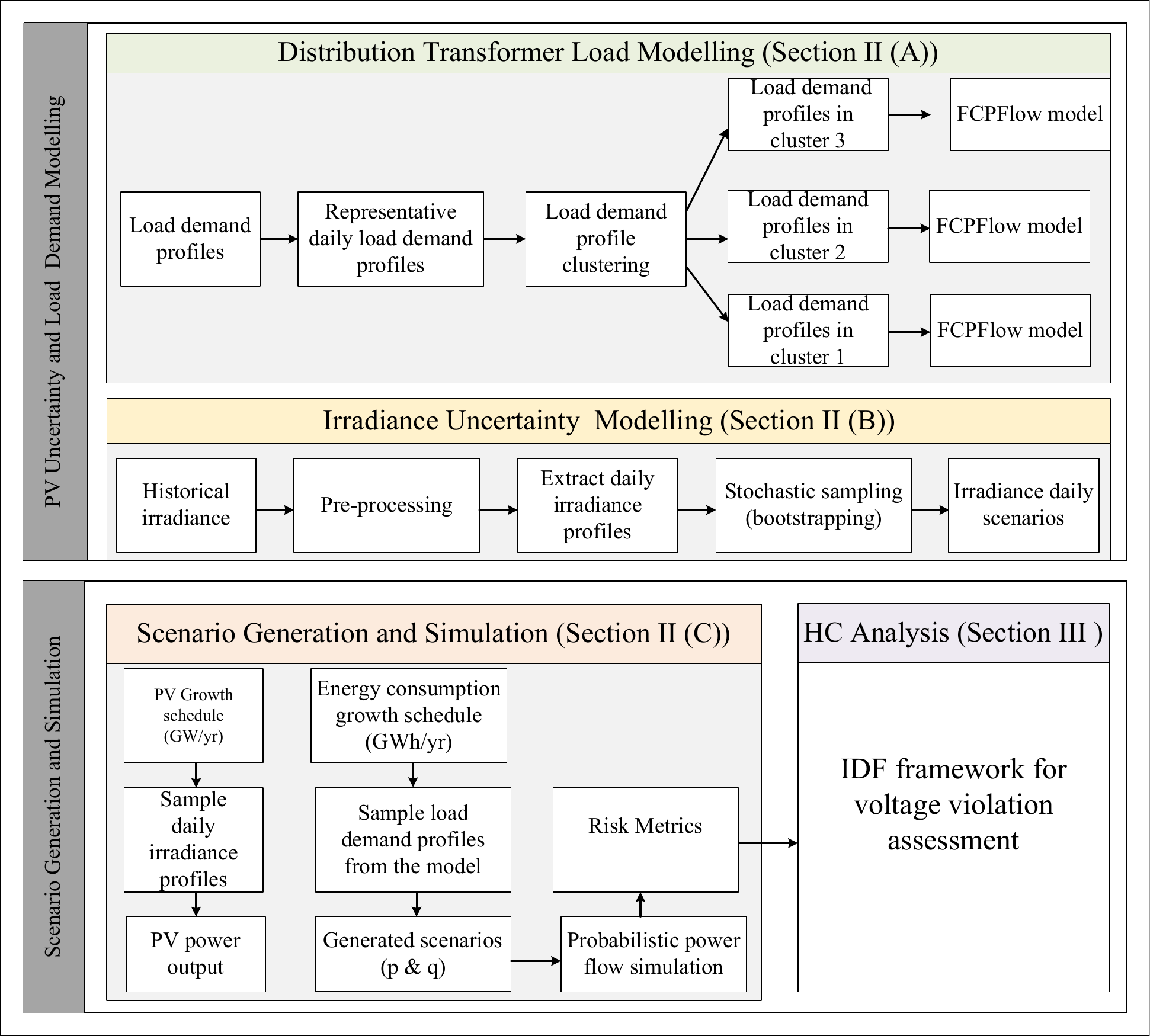}
   \caption{The proposed framework includes four main components: distribution transformer load modeling, irradiance uncertainty modeling, scenario generation and probabilistic power flow simulation, and risk metrics for HC assessment.}
    \label{fig:framework}
\end{figure}

Fig.~\ref{fig:framework} illustrates the overall workflow of the proposed framework, and each of its components: distribution transformer load modeling, irradiance uncertainty modeling, scenario generation and simulation, and risk metrics for HC assessment. Next, a detailed description of each component is presented.

\subsection{Distribution Transformer Load Modeling}
This section introduces the deep generative model used to generate load demand profiles of MV distribution transformers conditioned on their annual energy consumption.
\subsubsection{Transformer Clustering}
We first cluster distribution transformers based on their historical load demand profiles to capture similarities in consumption behavior across the distribution system. Clustering transformers into specific groups enables us to develop cluster-specific generative models that represent residential, commercial, industrial, and mixed demand patterns in the serviced area. 
Let $M$ denote the total number of MV distribution transformers, 
and $N$ denotes the total number of records over the observation period. For each transformer $m \in \{1,\dots,M\}$, a historical load demand profile is available, 
consisting of measurements recorded at uniform time steps. 
The load demand profile of transformer $m$ is represented as a discrete-time sequence
\vspace{-6pt}
\begin{equation}
\mathbf{p}_{m}^{(n)} = \left[ P_{m,1}^{(n)}, P_{m,2}^{(n)}, \dots, P_{m,T}^{(n)} \right] \in \mathbb{R}^{T}
\label{eq:transformer_load_profile}
\end{equation}
where $P_{m,t}^{(n)}$ denotes the active power demand at time step $t$ of the $n$-th load demand profile, 
with $n \in \{1,\dots,N\}$ and $t \in \{1,\dots,T\}$. 
Here, $T$ represents the number of time steps per profile 
(e.g., $T=96$ for daily profiles at 15-minute resolution).

To obtain a representative characterization of the typical daily loading behavior of each transformer, a representative load demand profile is constructed by averaging the available profile realizations given in \eqref{eq:transformer_load_profile} at each time step. Specifically, for transformer $m$, it is defined as
\vspace{-6pt}
\begin{small}
\begin{equation}
\bar{\mathbf{p}}_m =
\left[ \bar{P}_{m,1}, \bar{P}_{m,2}, \dots, \bar{P}_{m,T} \right] \in \mathbb{R}^{1 \times T},
\label{eq:rlp_definition}
\end{equation}
where
\vspace{-6pt}
\begin{equation}
\bar{P}_{m,t} = \frac{1}{N} \sum_{n=1}^{N} P_{m,t}^{(n)}, 
\quad t \in \{1,\dots,T\}.
\label{eq:rlp_mean}
\end{equation}
\end{small}

Then, a matrix, $\bar{\mathbf{P}} \in \mathbb{R}^{M \times T}$, is constructed by stacking the representative load demand profiles of all $M$ transformers, where each row corresponds to a transformer and each column represents a time step.
\begin{small}
\begin{equation}
\bar{\mathbf{P}} =
\begin{bmatrix}
\bar{\mathbf{p}}_1 \\
\bar{\mathbf{p}}_2 \\
\vdots \\
\bar{\mathbf{p}}_M
\end{bmatrix}
=
\begin{bmatrix}
\bar{P}_{1,1} & \cdots & \bar{P}_{1,T} \\
\bar{P}_{2,1} & \cdots & \bar{P}_{2,T} \\
\vdots        & \ddots & \vdots \\
\bar{P}_{M,1} & \cdots & \bar{P}_{M,T}
\end{bmatrix}
\in \mathbb{R}^{M \times T},
\label{eq:rlp_matrix}
\end{equation}
\end{small}

The clustering task is formulated as an unsupervised learning problem, as no prior labels are available regarding the underlying consumption categories or the appropriate number of transformer groups. Therefore, multiple clustering algorithms are evaluated using the matrix defined in \eqref{eq:rlp_matrix}. The optimal number of clusters and the most suitable clustering method are selected based on several internal validity metrics that assess cluster compactness and separation. A detailed comparison of the evaluated clustering methods and the corresponding performance metrics is provided in Section~III (A).

Let $c(m) \in \{1,\dots,K\}$ denotes the cluster assignment of transformer $m$, where $K$ is the selected number of clusters. The set of transformers belonging to the cluster $k$ can be defined as shown in \eqref{eq:cluster_set_definition}. These cluster-specific transformer sets are subsequently used to train separate generative models for each cluster group.
\begin{small}
\begin{equation}
\mathcal{M}_k = \{ m \in \{1,\dots,M\} \mid c(m) = k \}, 
\quad k \in \{1,\dots,K\}.
\label{eq:cluster_set_definition}
\end{equation}
\end{small}
\subsubsection{Conditional Flow-Based Generative Model}
A conditional flow-based deep generative model is used to generate load demand scenarios. The used model was originally proposed in \cite{flow_model} and learns the probability distribution of daily transformer active power profiles and generates synthetic load demand profiles conditioned on annual energy consumption. Reactive power is not modeled explicitly; instead, it is computed from the generated active power using a predefined power factor.  Further details on the model architecture, conditioning,  hyperparameter selection, and training mechanisms can be found in \cite{flow_model}.

\paragraph{Cluster-wise Conditional Modeling}

Let $w_{m}^{(n)} \in \mathbb{R}$ denote the annual energy consumption in GWh/year
associated with the $n$-th daily active power profile of transformer $m$. 
For each cluster $k$, the corresponding training dataset is defined as
\begin{equation}
\mathcal{D}_k =
\left\{ \left( \mathbf{p}_{m}^{(n)},\, w_{m}^{(n)} \right)
\;\middle|\;
m \in \mathcal{M}_k,\;
n \in \{1,\dots,N\} \right\}.
\label{eq:cluster_dataset}
\end{equation}

Based on the dataset defined in \eqref{eq:cluster_dataset}, a separate conditional flow-based generative model is trained for each cluster to capture the distinct consumption patterns associated with different user types.
(e.g., residential, industrial, commercial).

\paragraph {Conditional Generative Model}

For a given cluster $k$, the conditional probability density of daily active-power profiles, given annual energy consumption $ w$, can be modeled as
\vspace{-6pt}
\begin{equation}
p_{\theta,k}\!\left( \mathbf{p} \mid w \right),
\label{eq:cond_density}
\end{equation}
where $\mathbf{p} \in \mathbb{R}^{T}$ denotes the vectorized daily load demand profile and $w \in \mathbb{R}$.

A conditional normalizing flow model defines an invertible transformation
\vspace{-6pt}
\begin{equation}
\mathbf{p} = f_{\theta,k}(\mathbf{z}; w),
\qquad
\mathbf{z} \sim p(\mathbf{z}),
\label{eq:flow_forward}
\end{equation}
with inverse
\vspace{-6pt}
\begin{equation}
\mathbf{z} = f_{\theta,k}^{-1}(\mathbf{p}; w).
\label{eq:flow_inverse}
\end{equation}

Here, \( p(\mathbf{z}) \) denotes the base distribution, chosen as a standard multivariate Gaussian distribution, \( \mathcal{N}(\mathbf{0}, \mathbf{I}) \), 
and $f_{\theta,k}(\cdot;\,w)$ is an invertible neural network defined by the trainable parameters denoted by $\theta$. Using the Change of Variable Theorem\cite{papamakarios2021normalizing}, the conditional log-likelihood is expressed as
\vspace{-4pt}
\begin{equation}
\log p_{\theta,k}\!\left( \mathbf{p} \mid w \right)
=
\log p(\mathbf{z})
+
\log \left|
\det \left(
\frac{\partial f_{\theta,k}^{-1}(\mathbf{p}; w)}{\partial \mathbf{p}}
\right)
\right|,
\label{eq:change_of_variables}
\end{equation}
where $\mathbf{z} = f_{\theta,k}^{-1}(\mathbf{p}; w)$. \\

The model parameters $\theta$ are estimated by maximizing the conditional log-likelihood in \eqref{eq:change_of_variables} over the training dataset $\mathcal{D}_k$, as expressed in \eqref{eq:mle_training}. 
\vspace{-6pt}
\begin{equation}
\theta_k^\star =
\arg\max_{\theta}
\sum_{(\mathbf{p},w)\in\mathcal{D}_k}
\log p_{\theta,k}\!\left( \mathbf{p} \mid w \right).
\label{eq:mle_training}
\end{equation}

\subsection{Irradiance Uncertainty Modeling}

PV generation uncertainty is represented using a non-parametric, scenario-based approach constructed from historical irradiance measurements. This approach avoids distributional assumptions and preserves the empirical temporal variability observed in the data.

Let $G(t)$ denote the historical irradiance time series dataset. The data is segmented into daily profiles to capture intra-day variability while maintaining a realistic temporal structure. The resulting dataset is defined as
\vspace{-6pt}
\begin{equation}
\mathcal{G} = \left\{ \mathbf{g}_1, \mathbf{g}_2, \dots, \mathbf{g}_N \right\},
\end{equation}
where each daily profile is given by
\vspace{-6pt}
\begin{equation}
\mathbf{g}_d = \left[ G_{d,1}, G_{d,2}, \dots, G_{d,T} \right],
\end{equation}
and $T$ denotes the number of time steps per day. Irradiance scenarios are generated via bootstrap resampling with replacement from $\mathcal{G}$. For scenario $s$, the sampled profile satisfies
\vspace{-4pt}
\begin{equation}
\mathbf{g}^{(s)} \sim \text{Uniform}(\mathcal{G}),
\label{eq:pv_sample}
\end{equation}
ensuring that each historical day has an equal probability of selection while preserving realistic daily irradiance patterns.

\subsection{Scenario Generation and Simulation}

In distribution systems, the evolution of energy consumption over a defined planning horizon is inherently heterogeneous, as each MV transformer may exhibit a distinct growth trajectory. To represent this heterogeneity, transformers are grouped into clusters based on their demand characteristics, following the clustering approach described in Section~II-A. A dedicated generative model is then trained for each cluster, as defined in \eqref{eq:mle_training}. These cluster-specific generative models are subsequently used to generate load demand scenarios for different annual energy consumption levels over the planning horizon.

\begin{figure}[t]
    \centering
    \includegraphics[width=\linewidth]{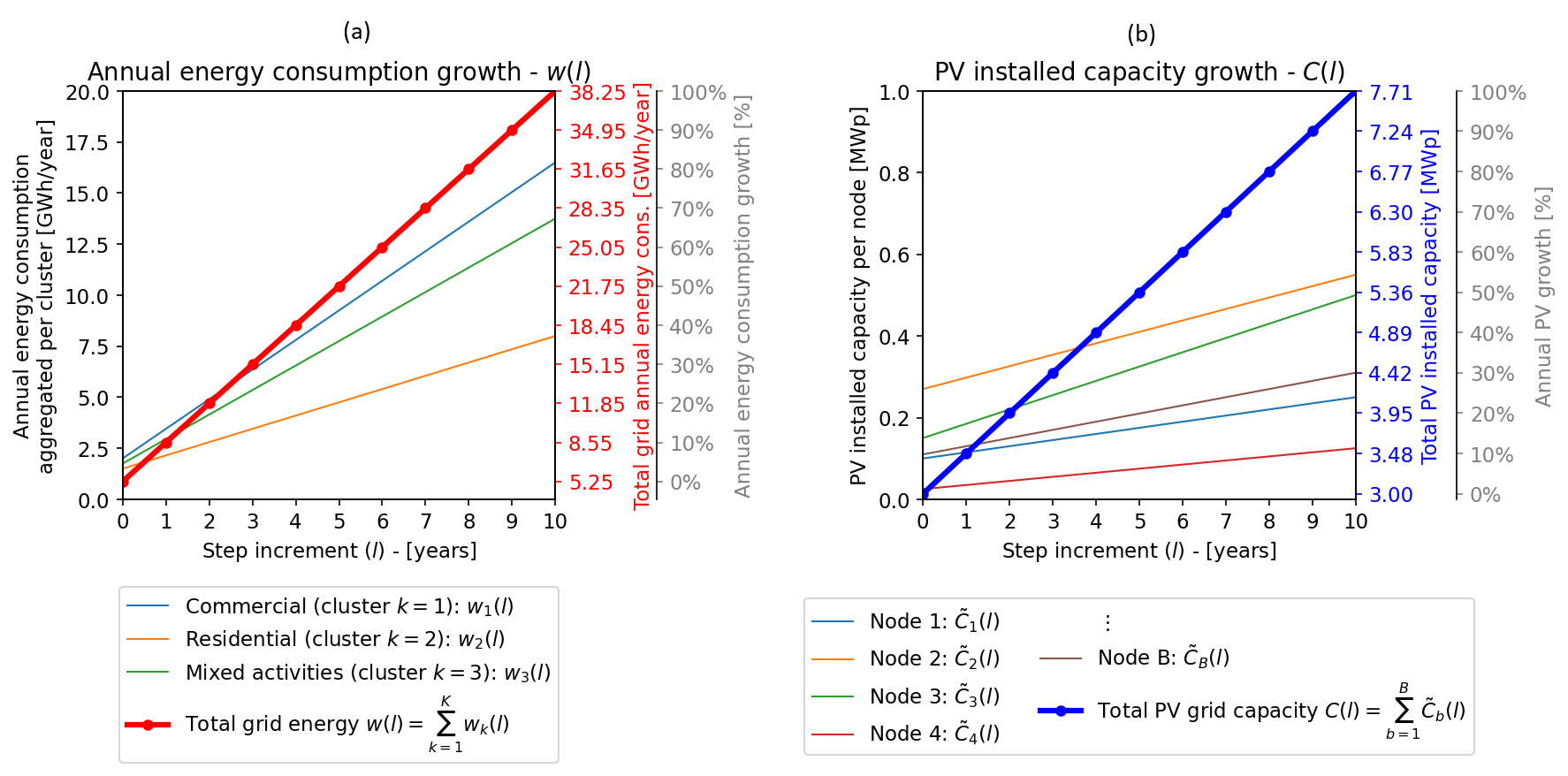}
\caption{Linear growth of (a) annual energy consumption and (b) PV installed capacity over the planning horizon. Solid lines represent cluster/node-level trajectories, while bold lines indicate total system growth. Percentages denote normalized scaling between minimum and maximum planning bounds.}
\label{fig:load_pv_growth}
\end{figure}

From a planning perspective, DSOs typically specify the expected annual energy consumption at the end of the planning horizon. Let $w_k^{\min}$ and $w_k^{\max}$ denote the base-year and maximum annual energy consumption (in GWh/year) of cluster $k$, respectively. Assuming a monotonically increasing growth trajectory, the annual energy consumption at each intermediate planning step $l$ is defined as
\begin{equation}
w_k(l) = w_k^{\min} + \gamma(l)\left( w_k^{\max} - w_k^{\min} \right),
\end{equation}
where $\gamma(l) \in [0,1]$ is a monotonically increasing growth function satisfying $\gamma(0)=0$ and $\gamma(L)=1$, with $L$ denoting the total number of planning steps. In this work, a linear growth trajectory is considered for illustration, i.e., $\gamma(l)=\frac{l}{L}$, as shown in Fig.~\ref{fig:load_pv_growth}. The total annual energy consumption of the grid at planning step $l$ is then obtained as
\begin{equation}
w(l) = \sum_{k=1}^{K} w_k(l),
\end{equation}
where $K$ is the number of transformer clusters.

Given a target annual energy consumption $w_k(l)$, synthetic daily load demand profiles are generated using the trained flow-based generative model for each cluster. For scenario generation, the model is conditioned on the annual energy level $w_k(l)$, and daily load demand scenarios are sampled as
\begin{equation}
\mathbf{z} \sim p(\mathbf{z}), \qquad
\hat{\mathbf{p}} = f_{\theta_k^\star,k}(\mathbf{z}; w_k(l)),
\label{eq:sampling}
\end{equation}
where $\hat{\mathbf{p}}$ represents a synthetic daily active power profile conditioned on the specified annual energy level. The corresponding reactive-power profile is obtained by assuming a constant power factor $\mathrm{pf}$, given by
\vspace{-6pt}
\begin{equation}
\mathbf{q}(l) = \hat{\mathbf{p}}(l)\tan\!\left(\arccos(\mathrm{pf})\right).
\end{equation}
The sampling procedure in \eqref{eq:sampling} enables the generation of time-correlated daily load demand profiles that are statistically consistent with the historical data of each cluster, while preserving the dependence on annual energy consumption.

In addition to annual energy consumption growth, the evolution of installed PV capacity is considered over the planning horizon. Each node may exhibit a distinct PV capacity growth trajectory, reflecting the spatially distributed nature of PV adoption. The total installed PV capacity of the grid at the planning step $l$ is defined as
\vspace{-6pt}
\begin{equation}
C(l) = \sum_{b=1}^{B} \tilde{C}_b(l),
\end{equation}
where $\tilde{C}_b(l)$ denotes the installed PV capacity at node $b$, and $B$ is the total number of nodes in the MV distribution system.

To capture the uncertainty of PV generation, irradiance scenarios sampled in Section~II-B are used. For each Monte Carlo realization $s$, the sampled irradiance profiles $\mathbf{g}^{(s)}$ are combined with the installed PV capacity at each node $\tilde{C}_b(l)$ to obtain the corresponding PV active power profiles $\mathbf{p}_b^{\text{pv},(s)}(l)$. This yields time-correlated PV generation profiles that capture both long-term capacity growth and irradiance variability.

For each planning step $l$ and MC realization $s$, the net active power profile at node $b$ is obtained as the difference between the generated load demand profiles and the corresponding PV output, i.e.,
\vspace{-6pt}
\begin{equation}
\mathbf{p}_b^{(s)}(l) = \hat{\mathbf{p}}_b^{(s)}(l) - \mathbf{p}_b^{\text{pv},(s)}(l),
\end{equation}
where $\hat{\mathbf{p}}_b^{(s)}(l)$ denotes the synthetic active power profile of transformers from the flow-based model, and $\mathbf{p}_b^{\text{pv},(s)}(l)$ represents the PV active power profile at node $b$. Thus, for a given node $b$, planning step $l$, and MC realization $s$, the pair $\left(\mathbf{p}_b^{(s)}(l), \mathbf{q}_b^{(s)}(l)\right)$ defines the active and reactive power profiles at node $b$. For probabilistic power flow simulation, a scenario is defined as the set of active and reactive power profiles across all nodes, i.e.,
\begin{equation}
s_l^{(s)} = \left\{ \mathbf{p}_b^{(s)}(l), \mathbf{q}_b^{(s)}(l) \right\}_{b \in \mathcal{B}}.
\label{eq:scenario}
\end{equation}
The set of all scenarios at planning step $l$ is given by
\begin{equation}
\mathcal{S}_l = \left\{ s_l^{(s)} \; \big| \; s = 1, \dots, S \right\}.
\end{equation}
\vspace{-6pt}
\subsection{Risk Metrics Formulation}

To evaluate planning risk under joint energy consumption and PV growth, three complementary dimensions are considered, inspired by the Intensity–Duration–Frequency (IDF) concept widely used in hydrological engineering \cite{IDF_water}, \cite{GIUDICIANNI2026134370}. In this framework, \emph{frequency} describes how often voltage magnitude violations occur across scenarios, \emph{intensity} characterizes the severity of voltage deviations, and \emph{duration} reflects the time scale over which sustained voltage deviations are evaluated.

For each planning step $l$ shown in Fig.~\ref{fig:load_pv_growth}, let $V_{b,t}^{(s)}(l)$ denote the voltage magnitude at node $b \in \mathcal{B}$, time step $t$, and MC realization $s$. Let $\overline{V}$ and $\underline{V}$ denote the maximum and minimum allowable voltage magnitudes, respectively. To jointly account for temporal persistence and spatial aggregation, a duration-constrained voltage statistic is defined for each scenario. For a given duration threshold $\tau$, let $n_\tau = \tau / \Delta t$. The overvoltage statistic is defined as
\vspace{-6pt}
\begin{equation}
\Phi_s^{\mathrm{ov}}(\tau,l)
=
\max_t
\left(
\frac{1}{n_\tau}
\sum_{i=0}^{n_\tau-1}
\max_{b \in \mathcal{B}} V_{b,t+i}^{(s)}(l)
\right),
\end{equation}
\vspace{-6pt}
and the corresponding undervoltage statistic is defined as
\begin{equation}
\Phi_s^{\mathrm{uv}}(\tau,l)
=
\min_t
\left(
\frac{1}{n_\tau}
\sum_{i=0}^{n_\tau-1}
\min_{b \in \mathcal{B}} V_{b,t+i}^{(s)}(l)
\right).
\end{equation}

These statistics provide a scalar representation of the worst sustained voltage behavior over a duration $\tau$ for each scenario.

\subsubsection*{1) Frequency Metric}

The frequency metric quantifies how often violations occur across scenarios. For the duration-constrained statistic, it is defined as
\vspace{-6pt}
\begin{equation}
F_\tau(l)
=
\mathbb{P}
\left(
\Phi_s^{\mathrm{ov}}(\tau,l) > \overline{V}
\right),
\end{equation}
\vspace{-6pt}
with empirical estimate
\vspace{-6pt}
\begin{equation}
F_\tau(l)
=
\frac{1}{S}
\sum_{s=1}^{S}
\mathbf{1}
\left(
\Phi_s^{\mathrm{ov}}(\tau,l) > \overline{V}
\right).
\end{equation}

A similar definition applies to undervoltage.

\subsubsection*{2) Intensity Metric}

The intensity metric characterizes the severity of sustained voltage stress across scenarios. It is defined through the quantile of the duration-constrained voltage statistic:
\vspace{-6pt}
\begin{equation}
I_q(\tau,l)
=
\mathcal{P}_q\!\left(\Phi_s^{\mathrm{ov}}(\tau,l)\right),
\end{equation}
where $\mathcal{P}_q(\cdot)$ denotes the $q$-th percentile across all scenarios. This metric quantifies how severe sustained voltage deviations can be under a given risk level.
\subsubsection*{3) Duration Representation}
The duration parameter $\tau$ defines the time scale over which sustained voltage behavior is evaluated, thereby enabling the distinction between short-lived extremes and persistent voltage violations. In this work, duration is interpreted as the persistence of voltage magnitude violations, i.e., the length of time over which violations remain sustained. To capture this persistence, the representative duration is defined as the largest window $\tau$ for which a selected percentile of the duration-constrained voltage statistic remains in violation of the corresponding limit. Specifically, for overvoltage,
\vspace{-6pt}
\begin{equation}
\tau^{\mathrm{ov}}_q(l)
=
\max \left\{ \tau \in \mathcal{T} \; \middle| \;
\mathcal{P}_{q}\!\left(\Phi_s^{\mathrm{ov}}(\tau,l)\right) \ge \overline{V}
\right\},
\end{equation}
\vspace{-6pt}
and for undervoltage,
\begin{equation}
\tau^{\mathrm{uv}}_q(l)
=
\max \left\{ \tau \in \mathcal{T} \; \middle| \;
\mathcal{P}_{q}\!\left(\Phi_s^{\mathrm{uv}}(\tau,l)\right) \le \underline{V}
\right\},
\end{equation}

where $\mathcal{P}_q(\cdot)$ denotes the $q$-th percentile across scenarios and $\mathcal{T}$ is the set of considered duration windows.
\section{Case Study and Discussions}

\begin{figure}[t]
\centering
 \includegraphics[width=0.7\linewidth]{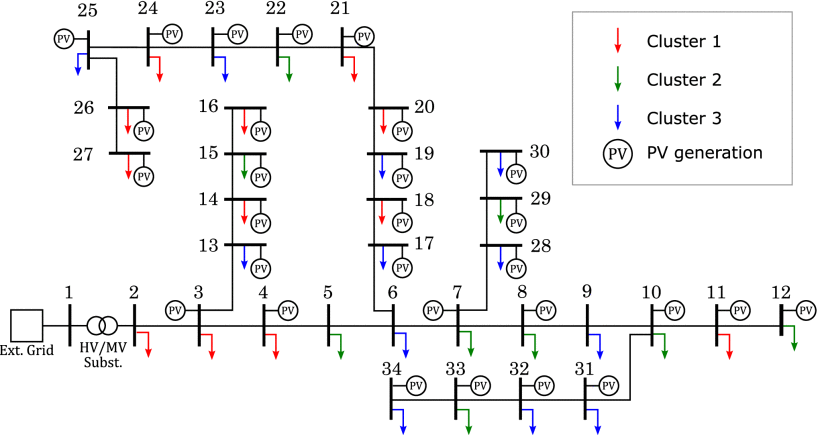}
 \caption{Radial MV distribution system used in the case study, showing PV installations and transformer clusters.}
 \label{fig:network_topology}
\end{figure}

The case study is based on the radial MV distribution system shown in Fig.~\ref{fig:network_topology}. The nominal substation voltage is set to 10~kV (1.0~p.u.), with a feeder thermal limit of 200~A. An upper operational voltage limit of 1.05~p.u. is adopted, while a caution threshold of 1.047~p.u. is introduced to identify operating conditions approaching voltage violations. To assess planning risk under uncertainty, probabilistic power flow simulations are performed for each planning step $l$. For every step, 1000 daily scenarios are generated to capture the stochastic variability of load demand and PV generation. Each scenario consists of 15-minute resolution active and reactive power time series defined in \eqref{eq:scenario}. In total, more than 11 million power flow calculations are carried out. 
\subsection{Transformer Clustering Results}

\begin{figure}[t]
\centering
 \includegraphics[width=0.92\linewidth]{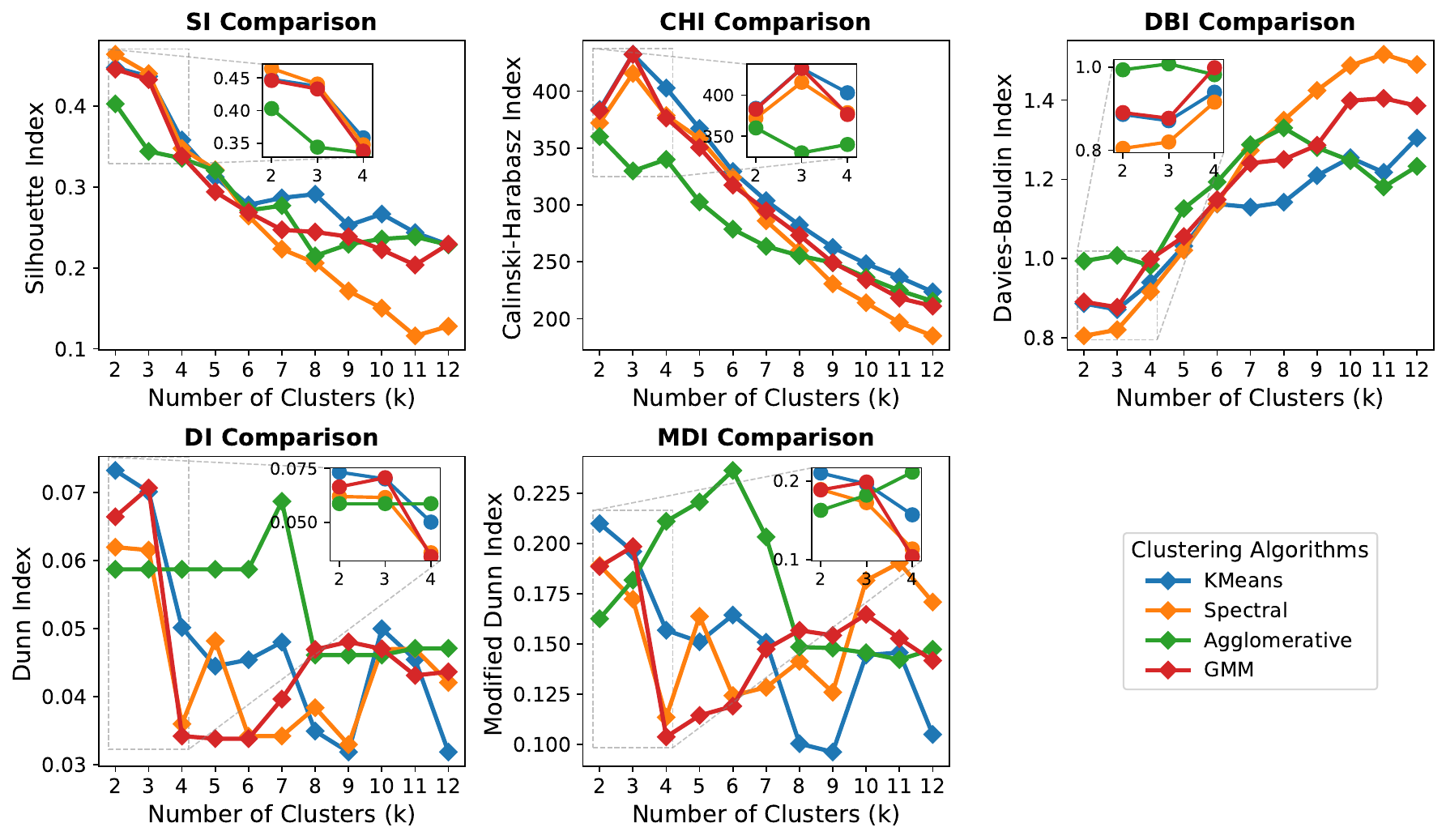}
 \caption{Clustering algorithm performance metrics for different cluster groups ($k=2$–$12$). Insets highlight $k=2$–$4$, where $k=3$ performs best.}
 \label{fig: clustering algorithms plot}
\end{figure}

The dataset consists of 476 MV transformers recorded at a 15-minute resolution. Four clustering algorithms, namely k-means, spectral, agglomerative, and Gaussian Mixture Models (GMM), are evaluated for $k=2$--$12$. Clustering performance is assessed using five validity indices: Silhouette Index (SI), Calinski–Harabasz Index (CHI), Davies–Bouldin Index (DBI), Dunn Index (DI), and Modified Dunn Index (MDI). Higher SI, CHI, DI, and MDI values indicate better clustering quality, whereas lower DBI values are preferred. Fig.~\ref{fig: clustering algorithms plot} presents the variation of these metrics with respect to $k$, with inset plots highlighting the range $k=2$--$4$. Across all algorithms, $k=3$ provides the most consistent trade-off between intra-cluster compactness and inter-cluster separation. Table~\ref{clustering_metrics} summarizes the results for $k=3$. Among the evaluated methods, GMM achieves the best overall performance, obtaining the highest values for CHI (432.44), DI (0.071), and MDI (0.199), while maintaining competitive SI and DBI values. For $k=3$, the GMM partitions the transformers into three groups: Cluster~1 with 272 transformers (57.1\%), Cluster~2 with 94 transformers (19.7\%), and Cluster~3 with 110 transformers (23.1\%). Based on these results, GMM is selected to cluster the available transformers.

\begin{table*}[t]
\centering
\renewcommand{\arraystretch}{1.2}
\caption{Summary Comparison of the clustering algorithms in terms of different metrics for $k=3$}
\label{clustering_metrics}
\scalebox{0.9}{
\begin{tabularx}{\textwidth}{
l
*{5}{>{\centering\arraybackslash}X}
>{\centering\arraybackslash}X
>{\centering\arraybackslash}X
>{\centering\arraybackslash}X
}

\toprule
\toprule
& \multicolumn{5}{c}{Clustering Metrics} & \multicolumn{3}{c}{Number of Transformers per Cluster} \\
\cmidrule(lr){2-6} \cmidrule(lr){7-9}
Algorithm & SI & CHI & DBI & DI & MDI & Cluster 1 & Cluster 2 & Cluster 3 \\
\midrule

KMeans
& \textbf{0.437139}
& 432.416326
& 0.871335
& 0.070122
& 0.196119
& 90 (18.9\%)
& 111 (23.3\%)
& 275 (57.8\%) \\

Spectral
& 0.40579
& 415.600897
& \textbf{0.820187}
& 0.061547
& 0.172371
& 302 (63.4\%)
& 101 (21.2\%)
& 73 (15.3\%) \\

Agglomerative
& 0.344376
& 329.916662
& 1.007992
& 0.058722
& 0.181894
& 169 (35.5\%)
& 107 (22.5\%)
& 200 (42.0\%) \\

GMM
& 0.433061
& \textbf{432.439510}
& 0.877219
& \textbf{0.070696}
& \textbf{0.198601}
& 272 (57.1\%)
& 94 (19.7\%)
& 110 (23.1\%) \\

\midrule
\bottomrule
\multicolumn{9}{l}{\footnotesize Best values: SI$\uparrow$, CHI$\uparrow$, DBI$\downarrow$, DI$\uparrow$, MDI$\uparrow$} \\

\bottomrule

\end{tabularx}
}
\vspace{1mm}
\label{clustering metrics}
\end{table*}

\subsection{Load Demand Profile Generation}

\begin{figure}[t]
    \centering
    \includegraphics[width=0.9\linewidth]{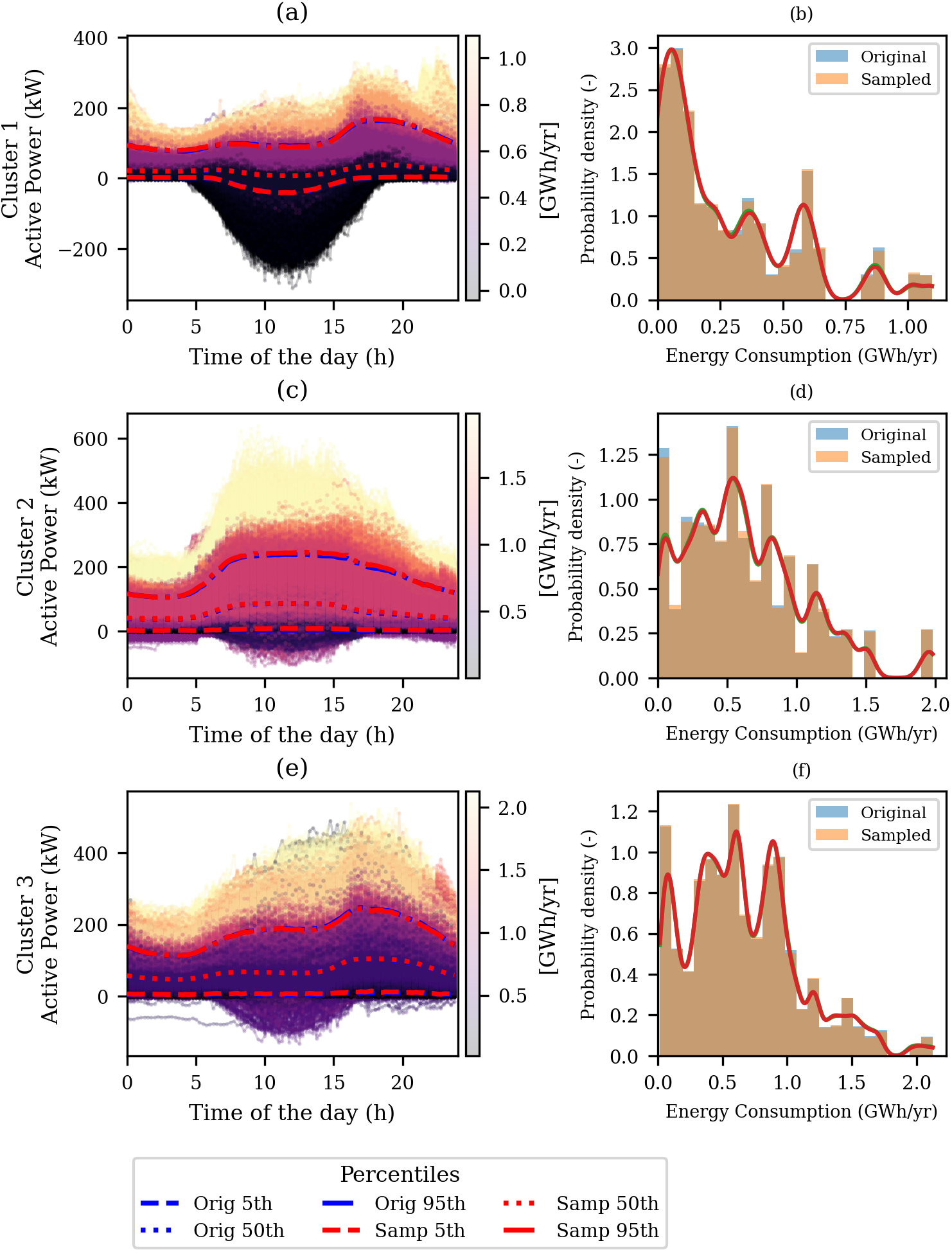}
\caption{Comparison of original and generated load demand profiles and annual energy distributions for three clusters. Panels (a), (c), and (e) show active power profiles with percentile bands, while (b), (d), and (f) show the corresponding PDFs.}

    \label{fig:all_clustering_profiles}
\end{figure}

Fig.~\ref{fig:all_clustering_profiles} compares the original and sampled daily load demand profiles for the three clusters together with their annual energy distributions. Distinct consumption behaviors are observed across the clusters. Cluster~1 exhibits a daytime plateau typical of commercial loads, Cluster~2 shows a pronounced evening peak associated with residential demand, while Cluster~3 presents a more mixed consumption pattern with midday reductions, likely influenced by PV penetration. For each cluster, the 5th, 50th, and 95th percentiles of the original and sampled profiles are shown. The close agreement between the percentile envelopes indicates that the generative models successfully preserve the temporal characteristics, variability, and peak behavior of the historical data. In addition, the probability density functions (PDFs) of annual energy consumption show strong agreement between the original and generated datasets, confirming that the conditioning mechanism preserves the targeted annual energy levels.

\begin{figure}[t]
    \centering
    \includegraphics[width=\linewidth]{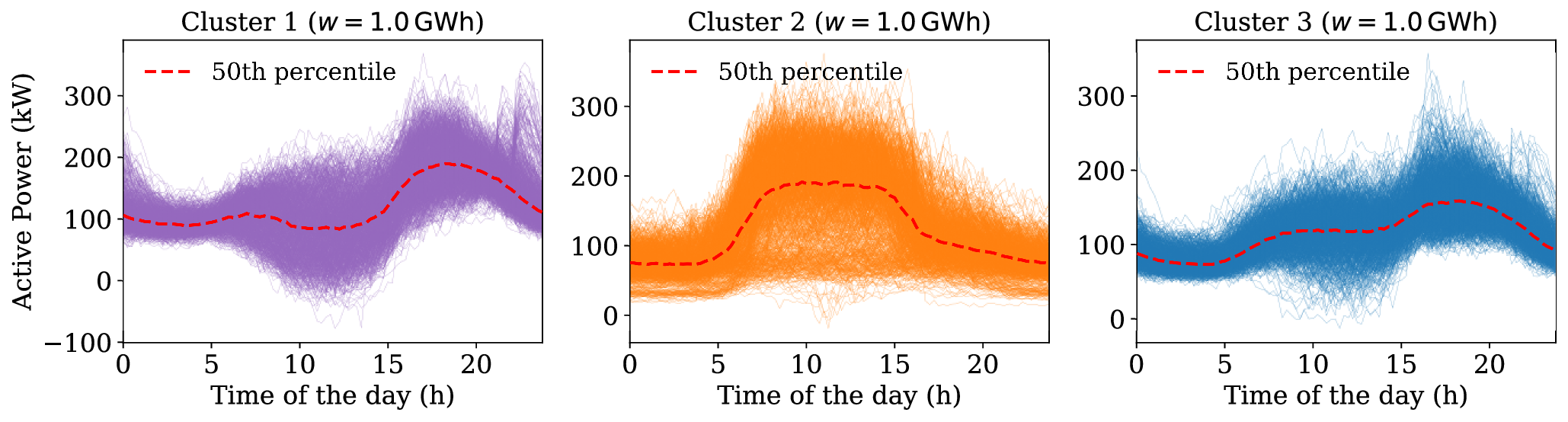}
    \caption{Generated daily load demand profiles for each cluster at an annual energy level of 1.0~GWh/year. The dashed line indicates the 50th percentile.}

    \label{fig: original and sampled load profiles}
\end{figure}

To further demonstrate the impact of clustering, additional experiments were performed by fixing the annual energy level to 1.0~GWh/year and generating multiple daily profiles for each cluster, as shown in Fig.~\ref{fig: original and sampled load profiles}. Although the annual energy is identical, the generated profiles exhibit distinct temporal patterns across clusters. This highlights the advantage of training separate cluster-specific generative models, which preserve the structural characteristics of different load types and avoid mixing heterogeneous consumption behaviors.

\subsection{Deterministic Voltage Magnitude Violation Assessment}

\begin{figure*}[t]
    \centering
    \includegraphics[width=0.85\textwidth]{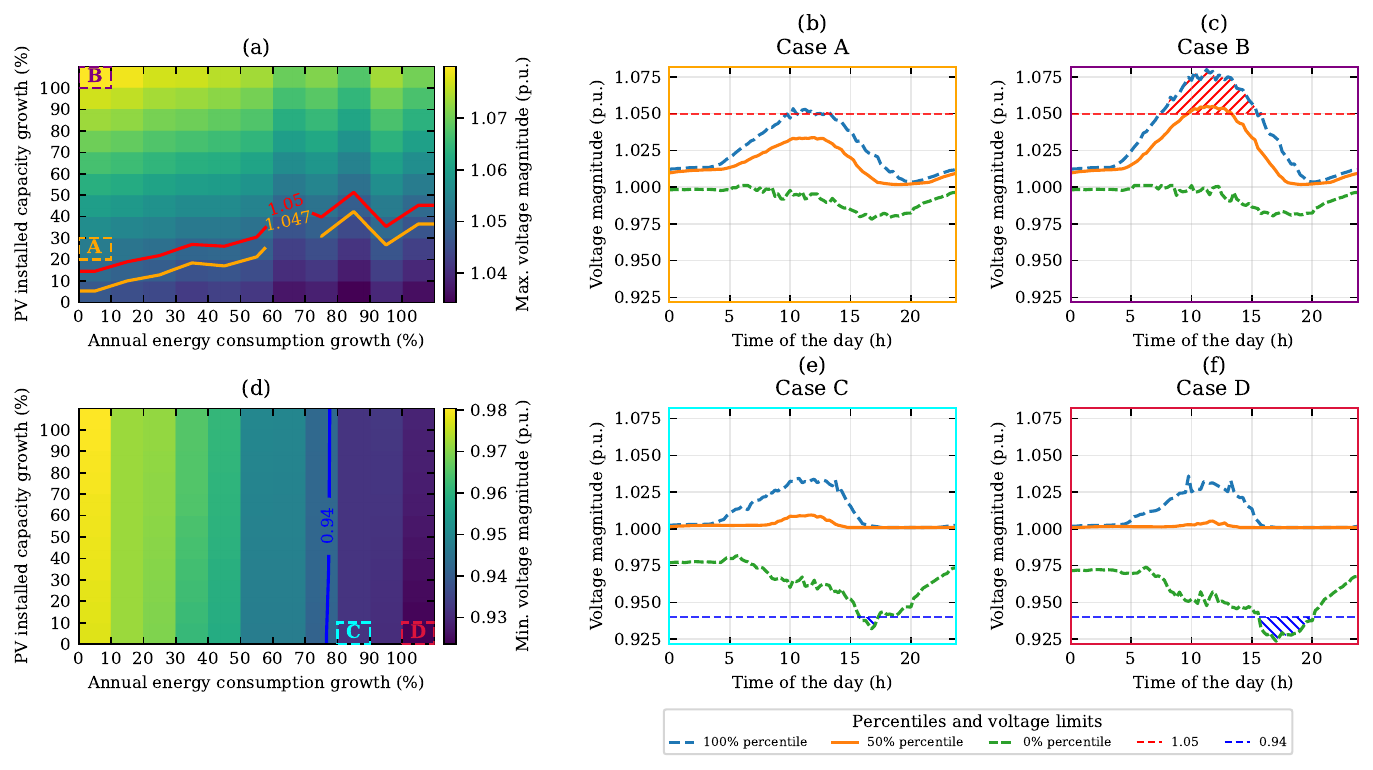}
    \caption{Voltage magnitude assessment based on extreme percentiles. 
(a), (d) Heatmaps for maximum and minimum voltage magnitudes across growth levels. 
(b)–(c) Voltage magnitude profiles for overvoltage cases; (e)–(f) voltage magnitude profiles for undervoltage cases. 
Shaded regions indicate voltage violations.}
    \label{fig:deterministic_voltage_assessment}
\end{figure*}

In this section, a deterministic (worst-case) assessment is performed using the extreme percentiles across all the simulated scenarios. Fig.~\ref{fig:deterministic_voltage_assessment} presents the voltage magnitude assessment across all energy consumption and PV growth combinations. Fig.~\ref{fig:deterministic_voltage_assessment}(a) and (d) show the maximum and minimum voltage magnitude heatmaps obtained from the 100th and 0th percentiles, respectively. The contour lines indicate the operational limits used for voltage magnitude violation assessment: 1.05~p.u. for overvoltage, 1.047~p.u. as a caution threshold, and 0.94~p.u. for undervoltage. It is worth noting that the undervoltage threshold is set to 0.94~p.u., slightly below the standard limit, to enhance visualization and better distinguish the impact of voltage magnitude violations across scenarios under deterministic assumptions.

\begin{figure}[t]
    \centering
    \includegraphics[width=0.65\columnwidth]{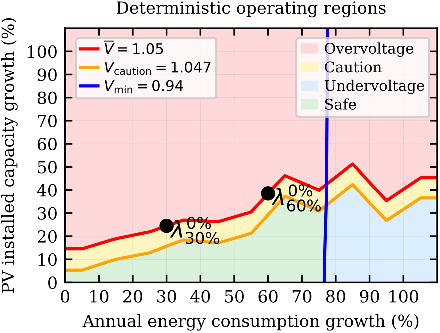}
    \caption{Deterministic voltage-based operating regions derived from extreme percentile assessment.     The red, orange, and blue lines indicate the overvoltage limit, caution threshold, and undervoltage limit, respectively, corresponding to a zero-risk (worst-case) planning perspective.}
    \label{fig:deterministic_operating_regions}
\end{figure}
Based on these extreme percentiles, the distribution system experiences voltage magnitude violations over a wide range of energy and PV growth combinations. As expected, overvoltage conditions dominate at high PV growth, whereas undervoltage conditions emerge at high energy growth levels. To further examine these conditions, four growth configurations are highlighted in Fig.~\ref{fig:deterministic_voltage_assessment}(a) and Fig.~\ref{fig:deterministic_voltage_assessment}(d) and analyzed through their corresponding percentile profiles in Fig.~\ref{fig:deterministic_voltage_assessment}(b) to Fig.~\ref{fig:deterministic_voltage_assessment}(f). Cases A $(0\%$ energy, $20\%$ PV) and B $(0\%$ energy, $100\%$ PV) illustrate overvoltage behavior driven by increasing PV growth. In contrast, Cases C $(80\%$ energy, $0\%$ PV) and D $(100\%$ energy, $0\%$ PV) capture undervoltage behavior associated with high energy growth. A key observation is that, when evaluated using extreme percentiles, all four cases are classified as violating operating limits, despite exhibiting significantly different temporal behaviors. 
\begin{table}[t]
\centering
\caption{Frequency, intensity, and duration metrics for the selected operating points (A--D).}
\label{tab:deterministic_quantification}
\scalebox{0.9}{
\begin{tabular}{lcccc}
\toprule
\toprule
 & \multicolumn{2}{c}{Overvoltage} & \multicolumn{2}{c}{Undervoltage} \\
 & A & B & C & D \\
\midrule
Frequency (\%)   & 1.64   & 96.44  & 1.37   & 15.89 \\
Intensity (p.u.) & 1.0534 & 1.0800 & 0.9321 & 0.9236 \\
Duration (h)     & 1      & 12     & 2      & 6 \\
\bottomrule
\bottomrule
\end{tabular}
}
\vspace*{0.5mm}

\end{table}
Nevertheless, the frequency, intensity, and duration of voltage magnitude violations vary substantially across the cases, as illustrated by the shaded regions in Fig.~\ref{fig:deterministic_voltage_assessment}(b)–(f) and quantified in Table~\ref{tab:deterministic_quantification}. For example, Case A exhibits only limited and short-duration overvoltage events, with a frequency of 1.64\% and a maximum duration of approximately 1~hour. In contrast, Case B shows severe and persistent violations, with a frequency exceeding 96\% and a duration of 12~hours. Similarly, the undervoltage cases demonstrate different levels of severity. Case C  exhibits infrequent violations (1.37\%) with short duration, whereas Case D experiences more frequent and sustained undervoltage conditions, with a frequency of 15.89\% and a duration of 6~hours. The intensity values further highlight the variation in violation severity across these operating points. For instance, Case A shows a mild violation with an intensity of 1.0534 p.u., whereas Case B reaches a significantly higher level of 1.08 p.u., indicating more severe overvoltage conditions. However, under a deterministic assessment based on the worst-case percentile, Cases A and B are treated equivalently, despite their fundamentally different violation characteristics. This highlights the inherent conservativeness of traditional worst-case analysis, where planning decisions are driven solely by extreme realizations without accounting for their likelihood or temporal duration. As a result, operating conditions characterized by infrequent or short-duration violations may be classified as infeasible, leading to overly conservative HC estimates and unnecessary grid reinforcement.
\begin{figure*}[!t]
    \centering
    \includegraphics[width=0.85\textwidth]{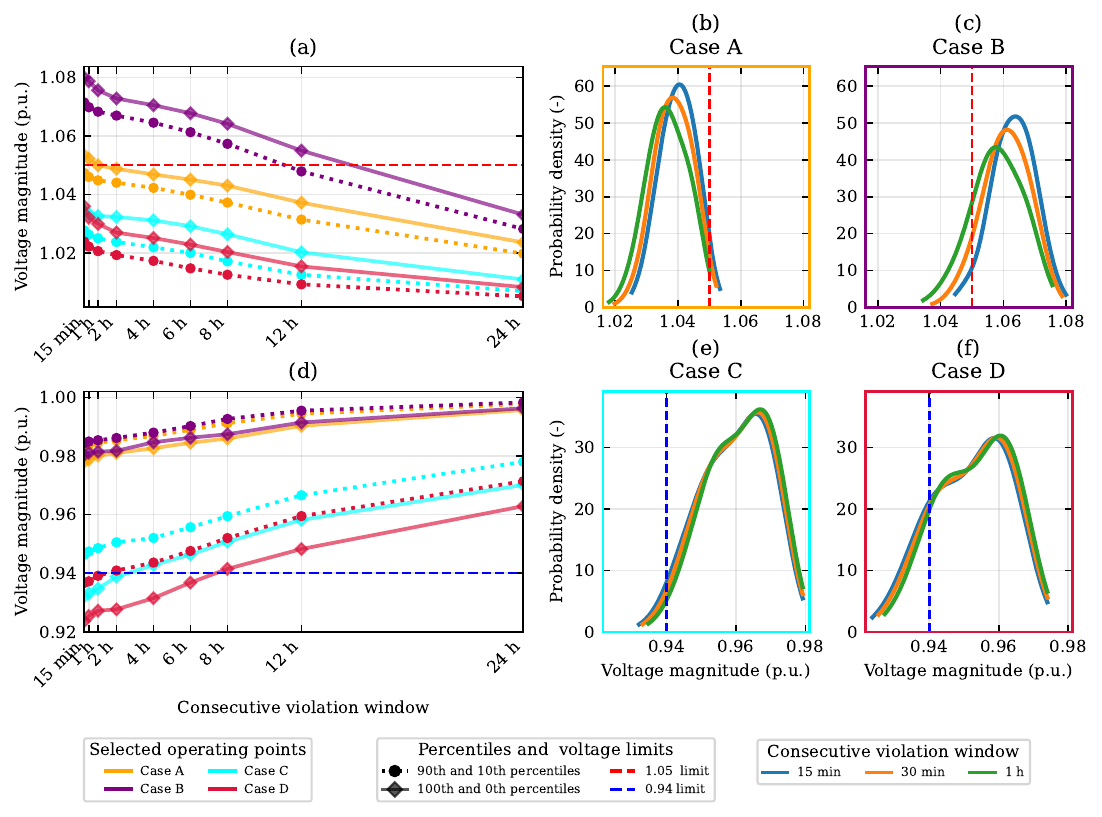}
\caption{Risk-based voltage violation assessment considering consecutive violation duration. 
(a), (d) Maximum and minimum voltage magnitudes versus violation window. 
(b)–(c) PDFs of maximum voltage (overvoltage cases); (e)–(f) PDFs of minimum voltage (undervoltage cases). 
Dashed lines indicate voltage limits.}
    \label{fig:voltage_violation_pdf}
\end{figure*}
The resulting operating regions, following the concept of static operating regions in \cite{risk_aware}, are shown in Fig.~\ref{fig:deterministic_operating_regions}. The energy and PV growth space is partitioned into safe, caution, overvoltage, and undervoltage regions corresponding to a zero-risk (worst-case) planning perspective. In Fig.~\ref{fig:deterministic_operating_regions}, $\lambda$ denotes an operating condition, where the subscript indicates the annual energy consumption growth level (in \%) and the superscript indicates the corresponding risk level (in \%). From these deterministic boundaries, the PV-HC can be directly quantified as the maximum PV growth level that does not exceed the overvoltage limit under the worst-case condition. For example, at $\lambda_{30\%}^{0\%}$ and $\lambda_{60\%}^{0\%}$, the corresponding HC is approximately 24.45\% and 38.5\%, respectively. While this representation provides a clear boundary of admissible operation, it does not distinguish between low-frequency and short-duration violations, highlighting the limitations of deterministic assessment and the need for more informative risk-based metrics
\subsection{Risk-Based Voltage Magnitude Violation Assessment}
\begin{figure*}[t]
    \centering
    \includegraphics[width=0.75\textwidth]{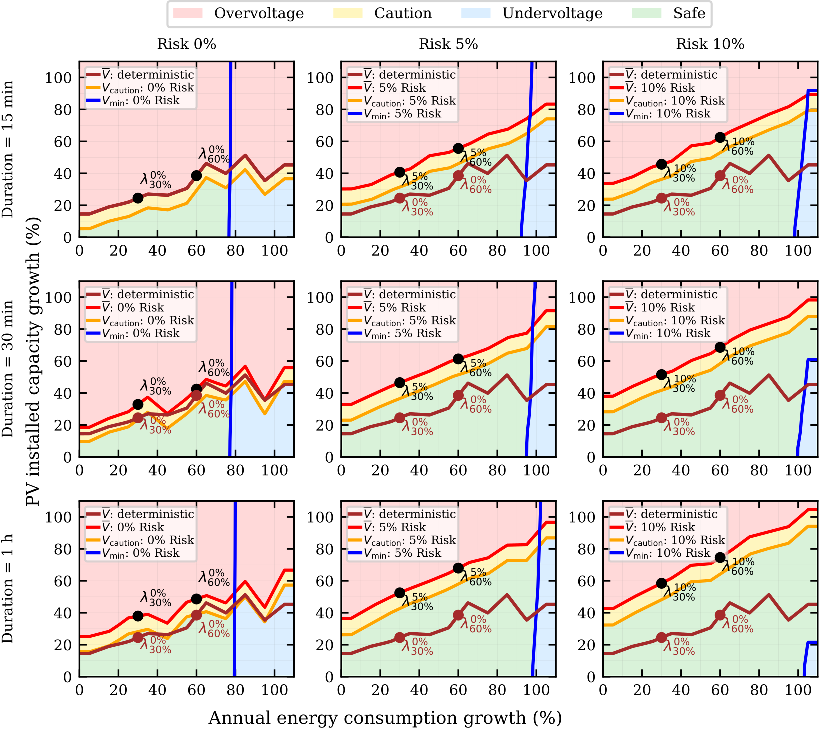}
    \caption{Risk-based operating regions for different risk levels and violation durations. 
Columns correspond to risk levels (0\%, 5\%, 10\%) and rows to duration thresholds (15~min, 30~min, 1~h). 
Contours indicate deterministic and risk-based voltage limits.}
    \label{fig:risk_based_operating_regions}
\end{figure*}
The proposed risk-based assessment builds upon the frequency, intensity, and duration metrics defined in Section~II and applies them within a probabilistic framework. For each scenario, voltage magnitude violations are evaluated over time, and the resulting metrics are aggregated across scenarios. The risk level (in \%) specifies the acceptable proportion of violating scenarios and is implemented through the corresponding percentile of the scenario-based results. For example, a 10\% risk level corresponds to evaluating the 90th percentile for overvoltage assessment. Fig.~\ref{fig:voltage_violation_pdf} illustrates how voltage magnitude varies with the consecutive violation window for selected growth configurations. The 90th percentile is used to assess overvoltage conditions, while the 10th percentile is used for undervoltage assessment, representing the acceptable risk level across scenarios. In addition, the 100th and 0th percentiles are included as deterministic references corresponding to extreme operating conditions. 

A clear trend observed in Fig.~\ref{fig:voltage_violation_pdf}(a) and (d) is that the intensity of voltage magnitude violations decreases as the allowed violation duration increases. This behavior can be illustrated using Case B (0\% energy growth and 100\% PV growth) in Fig.~\ref{fig:voltage_violation_pdf}(a). For a short-duration window of 15~min, the 90th percentile reaches 1.0715~p.u., while the corresponding extreme value reaches 1.0800~p.u. As the voltage violation duration window increases to 1~h, the 90th and 100th percentile values decrease to 1.0683~p.u. and 1.0756~p.u., respectively. For a 12~h window, the 90th percentile further decreases to 1.0479~p.u., falling below the 1.05~p.u. limit, whereas the deterministic extreme value remains slightly above the limit at 1.0550~p.u. A similar trend is observed for the undervoltage case. For Case D (100\% energy growth and 0\% PV growth), the 10th percentile of the minimum voltage is 0.9367~p.u. for a 15~min window and 0.9391~p.u. for a 1~h window, both below the 0.94~p.u. limit. As the allowed voltage violation duration window increases, the voltage rises above the minimum limit, reaching 0.9436~p.u. for a 4~h window, 0.9595~p.u. for a 12~h window, and 0.9714~p.u. for a 24~h window. In contrast, the deterministic extreme values remain significantly lower, ranging from 0.9236~p.u. for a 15~min window to 0.9629~p.u. for a 24~h window. The PDFs in Fig.~\ref{fig:voltage_violation_pdf}(b)--(f) further illustrate this behavior. For overvoltage, shorter window durations (15~min and 30~min) produce distributions shifted toward higher voltage magnitudes, with a large portion of the probability mass exceeding the 1.05~p.u. limit. This is confirmed by the corresponding violation probabilities: in Case B, 96.4\% of scenarios exceed the limit for a 15~min window, decreasing to 83.6\% for a 1~h window. A similar pattern is observed for Case D, where the probability mass below 0.94~p.u. decreases from 15.9\% for a 15~min window to 11.8\% for a 1~h window. Overall, based on these results, when we increase the allowed voltage violation windows, the estimated probability and intensity of voltage magnitude violations decrease. This due to the fact that short-lived voltage violations are smoothed out through moving-window averaging. As a result, we can increase the PV-HC, since many of these short-lived voltage violations can be managed during operation (e.g., through PV flexibility or curtailment). 
\begin{table}[t]
\centering
\caption{PV-HC (\%) under different risk levels and consecutive violation durations for selected energy growth levels.}
\label{tab:hc_summary}
\renewcommand{\arraystretch}{1.0}
\scalebox{0.9}{
\begin{tabular}{c c c c c}
\toprule
\toprule
Energy Growth & Duration & Risk 0\% & Risk 5\% & Risk 10\% \\
\midrule

30\% & 15 min & 24.45 & 40.70 & 45.65 \\
     & 30 min & 32.85 & 46.45 & 51.50 \\
     & 1 h    & 37.90 & 52.45 & 58.40 \\

\midrule

60\% & 15 min & 38.50 & 55.55 & 62.45 \\
     & 30 min & 42.45 & 61.30 & 68.60 \\
     & 1 h    & 48.60 & 67.95 & 74.55 \\

\bottomrule
\bottomrule
\end{tabular}
}
\end{table}

\begin{figure*}[t]
    \centering
    \includegraphics[width=0.82\textwidth]{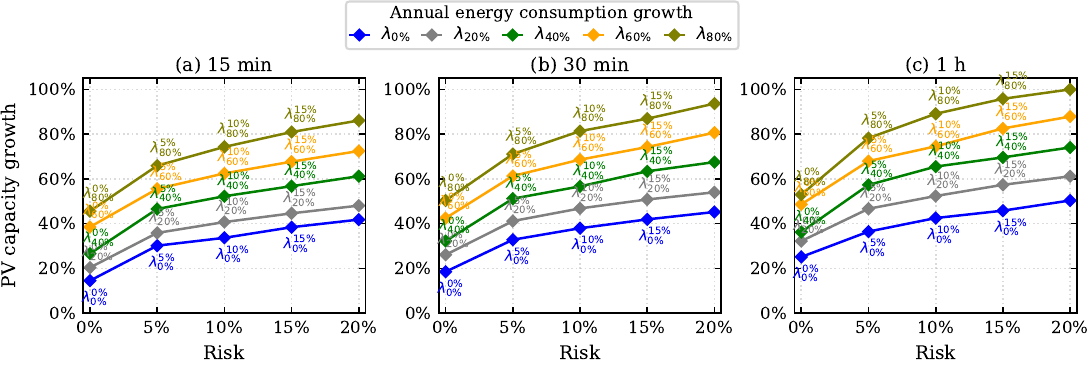}
    \caption{Quantification of PV-HC as a function of risk level, violation duration, and energy growth.
(a)–(c) correspond to durations of 15~min, 30~min, and 1~h. 
Curves represent different annual energy growth levels ($\lambda_{0\%}$–$\lambda_{80\%}$).}
    \label{fig:PV_HC_quantification}
\end{figure*}
Fig.~10 illustrates the impact of increasing the required duration and allowable risk levels through the duration-aware operating regions obtained using the proposed risk-based formulation. The columns correspond to risk levels of 0\%, 5\%, and 10\%, while the rows represent duration thresholds of 15~min, 30~min, and 1~h. The brown contour indicates the deterministic reference based on instantaneous (15~min) worst-case conditions. In Fig.~\ref{fig:risk_based_operating_regions}, a first observation is that the admissible operating region expands once a nonzero risk level is allowed. Compared to the deterministic reference, the 5\% and 10\% risk contours shift the overvoltage boundary upward, enabling higher PV integration. Second, the admissible region expands further as the required violation duration increases. This indicates that many voltage violations are short-lived and do not persist over longer time windows. Consequently, incorporating temporal persistence reclassifies several operating points from infeasible to acceptable.

Building upon the risk-based operating regions, the proposed framework also enables a direct quantification of PV-HC as a function of risk level, violation duration, and energy growth. Representative results for selected energy growth levels are summarized in Table~\ref{tab:hc_summary}. For instance, at 30\% energy growth, the PV-HC increases from 24.45\% under 0\% risk (and 15~min duration) to 40.70\% and 45.65\% for 5\% and 10\% risk, respectively. When the duration is increased further to 1~h, the corresponding PV-HC rises from 37.90\% to 52.45\% and 58.40\% for 5\% and 10\% risk levels, respectively. A similar trend is observed at 60\% energy growth, where the HC increases from 38.50\% (0\%, 15~min) to 55.55\% and 62.45\% for 5\% and 10\% risk, respectively, and further to 74.55\% for a 1~h duration. These results clearly demonstrate that both relaxing the risk constraint and allowing longer-duration violations lead to substantial increases in  PV-HC. Fig.~\ref{fig:PV_HC_quantification} further illustrates these trends across all energy growth levels. For a fixed energy growth, the admissible PV capacity increases monotonically with the allowed risk level, while for a fixed risk level, it also increases with the violation duration. This behavior indicates that a significant portion of voltage magnitude violations are short-lived. This also highlights the limitations of deterministic approaches, which are overly conservative as they treat all violations as equally critical regardless of their likelihood and temporal persistence. In contrast, the proposed framework provides a more flexible and realistic characterization of system limits by explicitly accounting for both the frequency and persistence of violations, thereby enabling improved PV-HC assessment and planning decisions.
\section{Conclusion}
This paper presented a risk-based framework for PV-HC assessment in distribution systems that explicitly accounts for the frequency, intensity, and duration of voltage magnitude violations under the combined effects of energy consumption and PV growth. A probabilistic scenario-based approach is used to capture realistic time-correlated operating conditions. The results show that conventional zero-risk assessments based on extreme percentiles are inherently conservative, as they neglect both the likelihood and temporal persistence of violations. In contrast, the proposed framework shows that many voltage violations are short-lived and occur with low probability, thereby expanding the admissible operating region and enabling higher HC estimates. For instance, allowing a 5\% risk level increases HC by approximately 18\% for a 15~min violation duration. However, it is important to note that allowing risk in the planning assessment does not imply that voltage violations will be allowed during operation. Instead, operational strategies such as flexibility from DERs, including PV disconnection, can be employed to ensure zero-risk operation.


\balance
\bibliographystyle{IEEEtran}
\bibliography{references}

\begin{thebibliography}{10}
\providecommand{\url}[1]{#1}
\csname url@samestyle\endcsname
\providecommand{\newblock}{\relax}
\providecommand{\bibinfo}[2]{#2}
\providecommand{\BIBentrySTDinterwordspacing}{\spaceskip=0pt\relax}
\providecommand{\BIBentryALTinterwordstretchfactor}{4}
\providecommand{\BIBentryALTinterwordspacing}{\spaceskip=\fontdimen2\font plus
\BIBentryALTinterwordstretchfactor\fontdimen3\font minus \fontdimen4\font\relax}
\providecommand{\BIBforeignlanguage}[2]{{%
\expandafter\ifx\csname l@#1\endcsname\relax
\typeout{** WARNING: IEEEtran.bst: No hyphenation pattern has been}%
\typeout{** loaded for the language `#1'. Using the pattern for}%
\typeout{** the default language instead.}%
\else
\language=\csname l@#1\endcsname
\fi
#2}}
\providecommand{\BIBdecl}{\relax}
\BIBdecl

\bibitem{DER}
H.~Lee, ``Integrating distributed energy resources in power distribution systems: A comprehensive review of impacts, challenges, and opportunities from a distribution system operator's perspective,'' \emph{Appl. Energy}, vol. 403, p. 127052, 2026.

\bibitem{DIPPENAAR2026102104}
J.~A. Dippenaar, B.~Bekker, K.~Foster, and M.~Davies, ``A synthesis of distributed energy resource impacts and regulatory responses with a focus on soft law,'' \emph{Utilities Policy}, vol.~98, p. 102104, 2026.

\bibitem{NPE2023}
\BIBentryALTinterwordspacing
{Ministry of Economic Affairs and Climate Policy}, ``Nationaal plan energiesysteem,'' 2023. [Online]. Available: \url{https://www.rijksoverheid.nl/documenten/rapporten/2023/12/01/nationaal-plan-energiesysteem}
\BIBentrySTDinterwordspacing

\bibitem{DER_review}
V.~de~Cillo~Moro, R.~S. Bonadia, and F.~C.~L. Trindade, ``A review of methods for assessing der hosting capacity of power distribution systems,'' \emph{IEEE Latin Amer. Trans.}, vol.~20, no.~10, pp. 2275--2287, 2022.

\bibitem{10415382}
H.~H.~H. Mousa, K.~Mahmoud, and M.~Lehtonen, ``A comprehensive review on recent developments of hosting capacity estimation and optimization for active distribution networks,'' \emph{IEEE Access}, vol.~12, pp. 18\,545--18\,593, 2024.

\bibitem{10811912}
A.~Benzerga, A.~Bahmanyar, G.~Derval, and D.~Ernst, ``A unified definition of hosting capacity, applications and review,'' \emph{IEEE Access}, pp. 1--1, 2024.

\bibitem{pedro_paper}
P.~P. Vergara, M.~Salazar, T.~T. Mai, P.~H. Nguyen, and H.~Slootweg, ``A comprehensive assessment of pv inverters operating with droop control for overvoltage mitigation in lv distribution networks,'' \emph{Renew. Energy}, vol. 159, pp. 172--183, 2020.

\bibitem{determistic}
A.~Dubey and S.~Santoso, ``On estimation and sensitivity analysis of distribution circuit's photovoltaic hosting capacity,'' \emph{IEEE Trans. Power Syst.}, vol.~32, no.~4, pp. 2779--2789, 2017.

\bibitem{determisitc2}
A.~Hoke, R.~Butler, J.~Hambrick, and B.~Kroposki, ``Steady-state analysis of maximum photovoltaic penetration levels on typical distribution feeders,'' \emph{IEEE Trans. Sustain. Energy}, vol.~4, pp. 350--357, 2013.

\bibitem{7913608}
H.~Al-Saadi, R.~Zivanovic, and S.~F. Al-Sarawi, ``Probabilistic hosting capacity for active distribution networks,'' \emph{IEEE Trans. Ind. Informat.}, vol.~13, no.~5, pp. 2519--2532, 2017.

\bibitem{8325320}
M.~S.~S. Abad, J.~Ma, D.~Zhang, A.~S. Ahmadyar, and H.~Marzooghi, ``Probabilistic assessment of hosting capacity in radial distribution systems,'' \emph{IEEE Trans. Sustain. Energy}, vol.~9, no.~4, pp. 1935--1947, 2018.

\bibitem{9745040}
C.~Han, D.~Lee, S.~Song, and G.~Jang, ``Probabilistic assessment of pv hosting capacity under coordinated voltage regulation in unbalanced active distribution networks,'' \emph{IEEE Access}, vol.~10, pp. 35\,578--35\,588, 2022.

\bibitem{ali2020probabilistic}
A.~Ali, K.~Mahmoud, D.~Raisz, and M.~Lehtonen, ``Probabilistic approach for hosting high pv penetration in distribution systems via optimal oversized inverter with watt-var functions,'' \emph{IEEE Syst. J.}, vol.~15, no.~1, pp. 684--693, 2020.

\bibitem{ma2024probabilistic}
H.~Ma, G.~Li, H.~Wang, Z.~Yan, and X.~Xu, ``Probabilistic evaluation of photovoltaic hosting capacity in unbalanced distribution network via polynomial chaos based kriging model,'' \emph{IEEE Trans. Ind. Appl.}, vol.~61, no.~1, pp. 1466--1474, 2024.

\bibitem{risk_aware}
E.~M.~S. Duque, J.~S. Giraldo, P.~P. Vergara, P.~H. Nguyen, A.~van~der Molen, and J.~G. Slootweg, ``Risk-aware operating regions for pv-rich distribution networks considering irradiance variability,'' \emph{IEEE Trans. Sustain. Energy}, vol.~14, no.~4, pp. 2092--2108, 2023.

\bibitem{10040563}
A.~N. Madavan, N.~Dahlin, S.~Bose, and L.~Tong, ``Risk-based hosting capacity analysis in distribution systems,'' \emph{IEEE Trans. Power Syst.}, vol.~39, no.~1, pp. 355--365, 2024.

\bibitem{9416873}
X.~Cao, T.~Cao, F.~Gao, and X.~Guan, ``Risk-averse storage planning for improving res hosting capacity under uncertain siting choices,'' \emph{IEEE Trans. Sustain. Energy}, vol.~12, no.~4, pp. 1984--1995, 2021.

\bibitem{flow_model}
W.~Xia, C.~Wang, P.~Palensky, and P.~P. Vergara, ``A flow-based model for conditional and probabilistic electricity consumption profile generation,'' \emph{Energy AI}, vol.~21, p. 100586, 2025.

\bibitem{papamakarios2021normalizing}
G.~Papamakarios, E.~Nalisnick, D.~Rezende, S.~Mohamed, and B.~Lakshminarayanan, ``Normalizing flows for probabilistic modeling and inference,'' \emph{J. Mach. Learn. Res.}, vol.~22, no.~57, pp. 1--64, 2021.

\bibitem{IDF_water}
D.~Koutsoyiannis, D.~Kozonis, and A.~Manetas, ``A mathematical framework for studying rainfall intensity-duration-frequency relationships,'' \emph{J. Hydrol.}, vol. 206, no.~1, pp. 118--135, 1998.

\bibitem{GIUDICIANNI2026134370}
C.~Giudicianni, F.~D. Nunno, F.~Granata, and E.~Creaco, ``Pseudo multi-scaling model of intensity--duration--frequency (idf) curves,'' \emph{J. Hydrol.}, vol. 664, p. 134370, 2026.

\end{thebibliography}

\end{document}